\documentstyle[aps, pra]{revtex}

\begin{document}
\title{Efficient quantum key distribution scheme with nonmaximally entangled states}
\author{Peng Xue, Chuan-Feng Li\thanks{%
Email address: cfli@ustc.edu.cn} , and Guang-Can Guo\thanks{%
Email address: gcguo@ustc.edu.cn}}
\address{Laboratory of Quantum Communication and Quantum Computation and \\
Department of Physics, University of Science and Technology of China,\\
Hefei 230026, P. R. China}
\maketitle

\begin{abstract}
\baselineskip12ptWe propose an efficient quantum key distribution scheme
based on entanglement. The sender chooses pairs of photons in one of the two
equivalent nonmaximally entangled states randomly, and sends a sequence of
photons from each pair to the receiver. They choose from the various bases
independently but with {\it substantially} different probabilities, thus
reducing the fraction of discarded data, and a significant gain in
efficiency is achieved. We then show that such a refined data analysis
guarantees the security of our scheme against a biased eavesdropping
strategy.

\end{abstract}

\baselineskip12pt

\section{Introduction}

Cryptography is the art of providing secure communication over insecure
communication channels. To achieve this goal, an algorithm is used to
combine a message with some additional information---known as the
``key''---to produce a cryptogram. For this reason, secure key distribution
is a crucial problem in cryptography.

Since the publication of BB84 scheme proposed by Bennett and Brassard, there
has been much interest in using quantum mechanics in cryptography \cite
{BB84,Ekert,BBM92,B92,BBBSS,Zeilinger,Kwiat,Gisin}. The security of these
quantum key distribution (QKD) schemes is based on the fundamental postulate
of quantum physics that ``every measurement perturbs a system''. Indeed,
passive monitoring of transmitted signals is strictly forbidden in quantum
mechanics. The ``quantum no-cloning theorem'' \cite{Dieks,Wootters}
indicates that it is impossible to make an exact copy of an unknown quantum
state.

Two well-known concepts for quantum key distribution are the BB84 scheme 
\cite{BB84} and the Ekert scheme \cite{Ekert}. The BB84 scheme \cite{BB84}
uses single photons transmitted between two parties (commonly called Alice
and Bob). The sender Alice uses non-orthogonal quantum states to transfer
the key to the receiver Bob. Such states cannot be cloned, hence any attempt
by an eavesdropper, known as Eve, to get information on the key disturbs the
transmitted signals and induces noise which will be detected during the
second stage of the transmission. Alice and Bob randomly pick a subset of
photons from those that are measured in correct bases and publicly compare
their measurements. For these results, they estimate the average error rate $%
\overline{e}$. If $\overline{e}$ turns out to be unreasonably large, then
eavesdropping has occurred, all the data are discarded and they may re-start
the whole procedure.

The Ekert scheme \cite{Ekert} is based on entangled pairs and uses the
generalized Bell's inequality (Clauser-Horne-Shimony-Holt inequality) \cite
{Bell,CHSH} to establish security. Both Alice and Bob receive one particle
out of an maximally entangled pair. They perform measurements along at least
three different directions on each side, where measurements along parallel
axes are used for key generation and oblique angles used for testing the
inequality

Neither scheme described above which is based on non-orthogonal quantum
cryptography has an efficiency more than $50\%$. Recently, Lo {\it et al}.
devise a modification \cite{Lo} that essentially doubles the efficiency of
the BB84 scheme, where, Alice and Bob choose between the two bases
independently but with {\it substantially} different probabilities $\epsilon 
$ and $1-\epsilon $. They also prove the security of their scheme.

In this paper, we present a new efficient QKD scheme with nonmaximally
entangled states.\ Suppose Alice creates pairs of photons in the
nonmaximally entangled state $\left| AB\right\rangle $ which can be
transformed to its equivalent state $\left| AB\right\rangle ^{\prime }$ with
the same Schmidt coefficients by local unitary transformations. She chooses
pairs of photons in one of the two states randomly, and sends a sequence of
photons out of each pair to Bob. The two users choose their bases
independently with different probabilities and perform measurements. Similar
to the scheme proposed by Lo {\it et al}., as two parties are much more
likely to be using the same basis, thus reducing the fraction of discarded
data, a significant gain in efficiency is achieved. To ensure our scheme is
secure, we separate the accepted data into various subsets according to the
basis employed and estimate an error rate for each subset {\it separately}.
We show that the refined error analysis is sufficient in ensuring the
security of our scheme against ``a biased eavesdropping attack'' \cite{Lo}.

In next section, we give the detailed description of our efficient QKD
scheme with nonmaximally entangled states. By considering a simple biased
eavesdropping strategy by Eve, we note that our refined analysis is an
essential feature of our scheme in Sec. III. In Sec. IV, the constraint on
the probability $\epsilon $ is derived. Finally, we conclude the scheme in
Sec. V.

\section{Efficient QKD scheme with nonmaximally entangled states}

In our scheme, there are two parties: the sender, Alice and the receiver,
Bob. Alice prepares pairs of photons in the nonmaximally entangled state 
\begin{equation}
\left| AB\right\rangle =\alpha \left| H\right\rangle _A\left| H\right\rangle
_B+\beta \left| V\right\rangle _A\left| V\right\rangle _B  \eqnum{1}
\end{equation}
where $\left| \alpha \right| ^2+\left| \beta \right| ^2=1$, and $H$ and $V$
denote the horizontal and vertical linear polarization, respectively. Then
she performs two ``$\sigma _x$'' operations on the two particles
respectively to transform the state $\left| AB\right\rangle $ to its
equivalent state 
\begin{equation}
\left| AB\right\rangle ^{\prime }=\beta \left| H\right\rangle _A\left|
H\right\rangle _B+\alpha \left| V\right\rangle _A\left| V\right\rangle _B 
\eqnum{2}
\end{equation}
with probability $\frac 12$. Photon B is sent to Bob and photon A is left
for Alice. There are two types of measurements that Alice may perform: she
may measure along the rectilinear basis, thus distinguishing between
horizontal and vertical photons. Alternatively, she may measure along the
diagonal basis, thus distinguishing between the $+45^o$ and $-45^o$ photons.
Bob measures the polarizations at the other end. He measures in one of three
bases, obtained by rotating rectilinear basis by angles $\phi _1\left( \phi
_1^{\prime }\right) =0$, $\phi _2\left( \phi _3^{\prime }\right) =\tan ^{-1}%
\frac \beta \alpha $, $\phi _3\left( \phi _2^{\prime }\right) =-\tan ^{-1}%
\frac \beta \alpha $. The surperscript ``$^{\prime }$'' refers to the case
in which Alice chooses $\left| AB\right\rangle ^{\prime }$ as the original
state.

The two users are connected by a quantum channel and a classical public
channel. The quantum channel consists usually of an optical fiber. The
public channel, however, can be any communication link. So how does this
scheme work?

1. Alice and Bob pick a number $0<\epsilon \leq 1$ and make its value
public. The constraint on $\epsilon $ will be discussed in Sec. IV.

2. Alice sends a sequence of photons B from each pair in one of the two
nonmaximally entangled states ($\left| AB\right\rangle $ and $\left|
AB\right\rangle ^{\prime }$) chosen randomly and independently, and leaves
the corresponding photons A. She also records her choice of $\left|
AB\right\rangle $ or $\left| AB\right\rangle ^{\prime }$.

3. Alice has two types of measurements. One measurement along rectilinear
basis (i.e., $\left\{ \left| H\right\rangle ,\left| V\right\rangle \right\} $%
) allows her to distinguish between horizontally and vertically polarized
photons. The other measurement along diagonal basis (i.e., $\left\{ \frac 1{%
\sqrt{2}}\left( \left| H\right\rangle +\left| V\right\rangle \right) ,\frac 1%
{\sqrt{2}}\left( \left| H\right\rangle -\left| V\right\rangle \right)
\right\} $) allows her to distinguish between photons polarized at $+45^o$
and $-45^o$. Alice chooses between the two types with probabilities $%
1-\epsilon $ and $\epsilon $, respectively. If she detects photon A in the
state $\left| H\right\rangle $ or $\frac 1{\sqrt{2}}\left( \left|
H\right\rangle +\left| V\right\rangle \right) $, the result is $0$; else,
the measurement can yield the result $1$, and potentially reveal one bit of
information. She writes down her measurement bases and the results of the
measurements.

4. For each photon, Bob performs measurements and registers the outcome of
the measurements in one of three bases, obtained by rotating the rectilinear
basis by angles $\phi _1\left( \phi _1^{\prime }\right) =0$, $\phi _2\left(
\phi _3^{\prime }\right) =\tan ^{-1}\frac \beta \alpha $, $\phi _3\left(
\phi _2^{\prime }\right) =-\tan ^{-1}\frac \beta \alpha $, i.e., $\left\{
\left| H\right\rangle ,\left| V\right\rangle \right\} $, $\left\{ \alpha
\left| H\right\rangle +\beta \left| V\right\rangle ,\beta \left|
H\right\rangle -\alpha \left| V\right\rangle \right\} $, and $\left\{ \beta
\left| H\right\rangle +\alpha \left| V\right\rangle ,\alpha \left|
H\right\rangle -\beta \left| V\right\rangle \right\} $, with probabilities $%
1-\epsilon $, $\frac \epsilon 2$, and $\frac \epsilon 2$, respectively.
Similar to the measurements performed by Alice, each measurement can yield
two results $0$ (if he detects photon B in the state $\left| H\right\rangle $%
, $\alpha \left| H\right\rangle +\beta \left| V\right\rangle $ or $\beta
\left| H\right\rangle +\alpha \left| V\right\rangle $) and $1$ (if he
detects photon B in the state $\left| V\right\rangle $, $\beta \left|
H\right\rangle -\alpha \left| V\right\rangle $ or $\alpha \left|
H\right\rangle -\beta \left| V\right\rangle $). The ensemble of these bits
registered by both Alice and Bob is the raw key.

5. After exchanging enough photons, Bob announces on the public channel the
sequence of bases he used, but not the results that he obtained.

6. Alice compares this sequence with the states that she originally chose,
and the list of polarizations which she measured. Then she tells Bob on the
public channel on which occasions his measurements have been done in the
correct bases. Whenever Alice and Bob used the compatible basis, they should
get perfectly correlated bits. However, due to imperfections in the setup,
and to a potential eavesdropper, there will be some errors.

There are two cases in which Alice chooses entangled states $\left|
AB\right\rangle $ and $\left| AB\right\rangle ^{\prime }$, respectively. For
either of the two cases, both Alice and Bob are much more likely to choose
the rectilinear basis and obtain correlated bits, thus achieving a
significant gain in efficiency. If Alice chooses the diagonal basis, in
order to generate a sifted key, Bob should choose between the bases $\left\{
\alpha \left| H\right\rangle +\beta \left| V\right\rangle ,\beta \left|
H\right\rangle -\alpha \left| V\right\rangle \right\} $, and $\left\{ \beta
\left| H\right\rangle +\alpha \left| V\right\rangle ,\alpha \left|
H\right\rangle -\beta \left| V\right\rangle \right\} $ according to the
entangled state chosen by Alice and the polarization of photon A.
(Otherwise, if he uses the rectilinear basis, he gets the outcomes $0$ and $%
1 $ with probabilities $\left| \alpha \right| ^2$ and $\left| \beta \right|
^2$, respectively. These results abort.) Foe example, if Alice chooses the
state $\left| AB\right\rangle $ and sends photon B to Bob. Then if she
detects photon A polarized at $+45^o$ by measuring along the diagonal basis,
Bob must choose the basis $\left\{ \alpha \left| H\right\rangle +\beta
\left| V\right\rangle ,\beta \left| H\right\rangle -\alpha \left|
V\right\rangle \right\} $ and photon B would be detected in the state $%
\alpha \left| H\right\rangle +\beta \left| V\right\rangle $. Therefore, they
can generate a key bit $``1"$ with probability $\frac 12\cdot \epsilon \cdot 
\frac \epsilon 2$. The bases used by Alice and Bob agree with probability $%
\left( 1-\epsilon \right) ^2+\frac{\epsilon ^2}2$ which goes to $1$ as $%
\epsilon $ goes to {\it zero.}

\begin{quote}
Table I. Example of the case where Alice chooses $\left| AB\right\rangle $
as the original state. The measurement bases are presented as the angles by
which the rectilinear basis is rotated (here $\theta =\tan ^{-1}\frac \beta %
\alpha $). The two users choose a basis with certain probability to measure
their particles and register the bit value ($0$ or $1$), respectively. The
ensemble of these bits is the raw key. Alice tells Bob on the public channel
on which occasions his measurements have been done in the correct bases, and
they keep only the bits corresponding to the compatible bases. This is the
sifted key.
\end{quote}

\begin{center}
\begin{tabular}{|c|c|c|c|c|c|c|c|c|c|c|c|c|}
\hline
A basis & $0$ & $0$ & $0$ & $0$ & $0$ & $0$ & $\frac \pi 4$ & $\frac \pi 4$
& $\frac \pi 4$ & $\frac \pi 4$ & $\frac \pi 4$ & $\frac \pi 4$ \\ \hline
A bit value & $0$ & $0$ & $0$ & $1$ & $1$ & $1$ & $0$ & $0$ & $0$ & $1$ & $1$
& $1$ \\ \hline
B basis & $0$ & $\theta $ & $-\theta $ & $0$ & $\theta $ & $-\theta $ & $0$
& $\theta $ & $-\theta $ & $0$ & $\theta $ & $-\theta $ \\ \hline
B bit value & $0$ & $1/0$ & $1/0$ & $1$ & $1/0$ & $1/0$ & $1/0$ & $0$ & $1/0$
& $1/0$ & $1/0$ & $1$ \\ \hline
compatible? & y & n & n & y & n & n & n & y & n & n & n & y \\ \hline
sifted key & $0$ &  &  & $1$ &  &  &  & $0$ &  &  &  & $1$ \\ \hline
\end{tabular}
\end{center}

7. For each of the two cases in which Alice chooses the entangled states $%
\left| AB\right\rangle $ or $\left| AB\right\rangle ^{\prime }$, Alice and
Bob divide up their polarization data into twelve cases according to the
actual bases used and the bit values yielded (shown in Table I),
respectively. Then they throw away the eight cases when they have used
non-compatible basis. Since{\it \ }the total probabilities for the two users
to obtain the results $0$ and $1$ are equal, the ensemble of these bits of
the remaining four cases is a sifted key. Hence, the remaining cases are
kept for further analysis and to generate the secret key.

8. Alice and Bob divide up the accepted data into two subsets according to
the entangled states originally chosen by Alice. From the subset where Alice
chooses $\left| AB\right\rangle $ as the prior state, there are three cases.
In one case where Alice and Bob both use the rectilinear basis (including
two cases shown in Table I, in each of which the bit value is $``0"$ or $%
``1" $), they randomly pick a fixed number say $m_1$ photons and publicly
compare their polarizations. The number of mismatches $r_1$ (here, mismatch
means the polarizations of photons are not correlated) tells them the
estimated error rate $e_1=\frac{r_1}{m_1}$. In the case where Alice uses the
diagonal basis and Bob uses the basis $\left\{ \alpha \left| H\right\rangle
+\beta \left| V\right\rangle ,\beta \left| H\right\rangle -\alpha \left|
V\right\rangle \right\} $, they pick a fixed number say $m_2$ photons and
publicly compare their polarizations. The number of mismatches $r_2$ gives
the estimated error rate $e_2=\frac{r_2}{m_2}$. In the case where Alice uses
the diagonal basis and Bob uses the basis $\left\{ \beta \left|
H\right\rangle +\alpha \left| V\right\rangle ,\alpha \left| H\right\rangle
-\beta \left| V\right\rangle \right\} $, they pick a fixed number say $m_3$
photons and publicly compare their polarizations. The number of mismatches $%
r_3$ gives the estimated error rate $e_3=\frac{r3}{m_3}$. Similarly, from
the subset where Alice chose $\left| AB\right\rangle ^{\prime }$ as the
prior state, there are also three cases. Corresponding to the above
discussion, we obtain the error rates $e_1^{\prime }=\frac{r_1^{\prime }}{%
m_1^{\prime }}$, $e_2^{\prime }=\frac{r_2^{\prime }}{m_2^{\prime }}$ and $%
e_3^{\prime }=\frac{r_3^{\prime }}{m_3^{\prime }}$.

Note that the test samples $m_1$, $m_1^{\prime }$, $m_2$, $m_2^{\prime }$, $%
m_3$ and $m_3^{\prime }$ are sufficiently large, the estimated error rates $%
e_1$, $e_1^{\prime }$, $e_2$, $e_2^{\prime }$, $e_3$, and $e_3^{\prime }$
should be rather accurate \cite{Chau,Shor}. Now they demand that $e_1$, $%
e_1^{\prime }$, $e_2$, $e_2^{\prime }$, $e_3$, and $e_3^{\prime }<e_{\max }$
where $e_{\max }$ is a prescribed maximal tolerable error rate. If these
independent constraints are satisfied, they proceed to the next steps.
Otherwise, they throw away the polarization data and re-start the whole
procedure. Notice that the constraints $e_1$, $e_1^{\prime }$, $e_2$, $%
e_2^{\prime }$, $e_3$, and $e_3^{\prime }<e_{\max }$ are more stringent than
the original naive prescription $\overline{e}<e_{\max }$ (here $\overline{e}$
is the average error rate). We will discuss it in detail in Sec. III.

9. Reconciliation and privacy amplification (see Ref. \cite{BB84,Lo}).

\section{Refined error analysis}

For each photon, the eavesdropper, Eve does not know which nonmaximally
entangled state it is chosen from. So for Eve, each photon is in an
entangled mixed state. She has eavesdropping attack as below:

i). with a probability $p_1$ measures polarization of each photon along the
rectilinear basis and re-sends a photon according to the result of her
measurement to Bob;

ii). with a probability $p_2$ measures polarization of each photon along the
basis $\left\{ \alpha \left| H\right\rangle +\beta \left| V\right\rangle
,\beta \left| H\right\rangle -\alpha \left| V\right\rangle \right\} $ and
re-sends a photon according to the result of her measurement to Bob;

iii). with a probability $p_3$ measures polarization of each photon along
the basis $\left\{ \beta \left| H\right\rangle +\alpha \left| V\right\rangle
,\alpha \left| H\right\rangle -\beta \left| V\right\rangle \right\} $ and
re-sends a photon according to the result of her measurement to Bob;

iv). with a probability $1-p_1-p_2-p_3$ does nothing.

Eve has a whole set of eavesdropping strategies by varying the values of $%
p_1 $, $p_2$ and $p_3$. Any of the strategies in this set is called ``a
biased eavesdropping attack'' \cite{Lo}.

Consider the error rate $e_1$ ($e_1^{\prime }$) for the case both Alice and
Bob use the rectilinear basis. For the biased eavesdropping strategy under
current consideration, errors occur only if Eve uses the other two bases.
This happens with a conditional probability $p_2+p_3$. In this case, the
polarization of the photon is randomized, thus giving an error rate 
\begin{equation}
e_1(e_1^{\prime })=2\alpha ^2\beta ^2\left( p_2+p_3\right) \text{.} 
\eqnum{3}
\end{equation}
Errors for the case where Alice uses the diagonal basis and Bob uses the
basis $\left\{ \alpha \left| H\right\rangle +\beta \left| V\right\rangle
,\beta \left| H\right\rangle -\alpha \left| V\right\rangle \right\} $ occur
only if Eve is measuring along the rectilinear basis or the basis $\left\{
\beta \left| H\right\rangle +\alpha \left| V\right\rangle ,\alpha \left|
H\right\rangle -\beta \left| V\right\rangle \right\} $. This happens with a
conditional probability $p_1+p_3$ and when it happens, the photon
polarization is randomized. Thus, the error rate for this case is 
\begin{equation}
e_2(e_3^{\prime })=2\alpha ^2\beta ^2p_1+8\alpha ^2\beta ^2\left( \alpha
^2-\beta ^2\right) ^2p_3\text{.}  \eqnum{4}
\end{equation}
Similarly, errors for the case where Alice uses the diagonal basis and Bob
uses the basis $\left\{ \beta \left| H\right\rangle +\alpha \left|
V\right\rangle ,\alpha \left| H\right\rangle -\beta \left| V\right\rangle
\right\} $ occur only if Eve is measuring along the rectilinear basis or the
basis $\left\{ \alpha \left| H\right\rangle +\beta \left| V\right\rangle
,\beta \left| H\right\rangle -\alpha \left| V\right\rangle \right\} $. This
happens with a conditional probability $p_1+p_2$. In this case, the error
rate is given as 
\begin{equation}
e_3(e_2^{\prime })=2\alpha ^2\beta ^2p_1+8\alpha ^2\beta ^2\left( \alpha
^2-\beta ^2\right) ^2p_2\text{.}  \eqnum{5}
\end{equation}
Therefore, Alice and Bob will find that, for the biased eavesdropping
attack, the average error rate 
\begin{eqnarray}
\overline{e} &=&\frac{\left( 1-\epsilon \right) ^2\left( e_1+e_1^{\prime
}\right) +\frac{\epsilon ^2}4\left( e_2+e_3+e_2^{\prime }+e_3^{\prime
}\right) }{2\left[ \left( 1-\epsilon \right) ^2+\frac{\epsilon ^2}2\right] }
\eqnum{6} \\
&=&\frac{\alpha ^2\beta ^2\left[ 2\left( 1-\epsilon \right) ^2\left(
p_2+p_3\right) +\epsilon ^2p_1+2\epsilon ^2\left( \alpha ^2-\beta ^2\right)
^2\left( p_2+p_3\right) \right] }{\left( 1-\epsilon \right) ^2+\frac{%
\epsilon ^2}2}\text{.}  \nonumber
\end{eqnarray}

Suppose Eve always eavesdrops only along rectilinear basis (i.e., $p_1=1$, $%
p_2=p_3=0$), then 
\begin{equation}
\overline{e}=\frac{\alpha ^2\beta ^2\epsilon ^2}{\left( 1-\epsilon \right)
^2+\frac{\epsilon ^2}2}\rightarrow 0  \eqnum{7}
\end{equation}
as $\epsilon $ tends to $0$, which is similar with the result of Ref. \cite
{Lo}. This means that if Eve is always eavesdropping along the dominant
basis, with a naive error analysis prescribed as $\overline{e}<e_{\max }$
Alice and Bob will fail to detect eavesdropping by Eve.

To ensure the security of our scheme, it is crucial to employ a refined data
analysis: the accepted data are further divided into various subsets
according to the actual basis used by Alice and Bob and the error rate of
each subset is computed separately. In Sec. II, we have already computed the
error rates $e_1$, $e_1^{\prime }$, $e_2$, $e_2^{\prime }$, $e_3$, and $%
e_3^{\prime }<e_{\max }$ where $e_{\max }$ is a prescribed maximal tolerable
error rate. From Eqs. (3,4,5), we can see that these error rates $e_1$, $%
e_1^{\prime }$, $e_2$, $e_2^{\prime }$, $e_3$, and $e_3^{\prime }$ depend on
Eve's eavesdropping strategy and the degree of entanglement of the original
state, but not on the value of $\epsilon $. So the refined data analysis
guarantees the security of the present scheme.

\section{The Constraint on $\epsilon $}

From the above discussion, we know the value of $\epsilon $ should be small
but can not be {\it zero}. If $\epsilon $ were actually {\it zero}, the
scheme would be insecure. The main constraint on $\epsilon $ is that there
should be enough photons for an accurate estimation of the six error rates $%
e_1$, $e_1^{\prime }$, $e_2$, $e_2^{\prime }$, $e_3$, and $e_3^{\prime }$.
We assume that $N$ entangled pairs are chosen by Alice, i.e., $N$ photons
are transmitted from Alice to Bob. On average, for $\left| AB\right\rangle $
or $\left| AB\right\rangle ^{\prime }$ only $N\epsilon ^2/8$ photons belongs
to each of the two cases where Alice uses the diagonal basis and Bob uses
the basis $\left\{ \alpha \left| H\right\rangle +\beta \left| V\right\rangle
,\beta \left| H\right\rangle -\alpha \left| V\right\rangle \right\} $ or the
basis $\left\{ \beta \left| H\right\rangle +\alpha \left| V\right\rangle
,\alpha \left| H\right\rangle -\beta \left| V\right\rangle \right\} $. To
estimate $e_2$, $e_2^{\prime }$, $e_3$, and $e_3^{\prime }$ reasonably
accurately, the number $N\epsilon ^2/8$ should be larger than some fixed
number say $m=\max \left( m_2,m_2^{\prime },m_3,m_3^{\prime }\right) $. The
numbers $m_2$, $m_2^{\prime }$, $m_3$, and $m_3^{\prime }$ are the photon
number needed for the refined error analysis, which can be computed from
classical statistical analysis. So 
\begin{eqnarray}
N\epsilon ^2/8 &\geq &m\text{,}  \eqnum{8} \\
\epsilon &\geq &2\sqrt{2m/N}\text{.}  \nonumber
\end{eqnarray}
As $N$ tends to $\infty $, $\epsilon $ can tend to {\it zero}, but never
reach it. And the efficiency of this scheme is asymptotic $100\%$.

\section{Discussion and Conclusion}

From the refined error analysis, we find the error rates depend not only on
Eve's eavesdropping strategy but also on the degree of entanglement of the
original state. For the biased eavesdropping attack, the error rates $e_i$
and $e_i^{\prime }$ ($i=1,2,3$) are functions of $\alpha \beta $, the
probability $\epsilon $ and the eavesdropping strategy of Eve (seeing Eqs.
(3-5)). If Alice uses a product state as the original state, i.e., $\alpha
\beta =0$, whatever the probability $\epsilon $ and Eve's eavesdropping
strategy are, the error rates $e_i$ and $e_i^{\prime }$ equal to {\it zero}.
That is, Alice and Bob will never detect eavesdropping by Eve whatever she
does. In other words, if $\alpha \beta =0$, the scheme is easily broken by
an eavesdropper. The security of our scheme is relying on the degree of the
entanglement of the original state. If $\left| \alpha \beta \right| =\frac 12
$, this scheme is equivalent to an efficient ``simplified EPR scheme'' \cite
{BBM92}.

Of course, the QKD with nonmaximally entangled states may also be completed
in another way. At first, the nonmaximally entangled state $\left|
AB\right\rangle =\alpha \left| 00\right\rangle +\beta \left| 11\right\rangle 
$ (here $\left| \beta \right| <\left| \alpha \right| $) can be concentrated
to an EPR state \cite{CH,Lo2} with probability $2\left| \beta \right| ^2$ 
\cite{Vidal1}. If the concentration fails, EPR pairs are abandoned; else,
they are used in an efficient ``simplified EPR scheme'' \cite{BBM92}.
Obviously, the total efficiency of this QKD process should be no more than $%
2\left| \beta \right| ^2$.

In summary, we propose a quantum key distribution (QKD) scheme based on
entanglement, where Alice and Bob choose between various bases independently
with substantially different probabilities. Since two parties are much more
likely to be using the same basis, thus reducing the fraction of discarded
data, a significant gain in efficiency is achieved. The efficiency can be
tend to $100\%$, as the value of $\epsilon $ tends to {\it zero} (but can
not reach it accurately).

To make the scheme secure against the dominant basis eavesdropping attack,
it is crucial to have a refined error analysis in place of a naive error
analysis. We separate the accepted data into various subsets according to
the basis employed and estimate an error rate for each subset separately. It
is only when all error rates are small enough that the security of
transmission is accepted.

\begin{center}
{\bf Acknowledgment}
\end{center}

This work was supported by the National Natural Science Foundation of China.

\end{document}